\documentclass[prb,aps,showpacs,superscriptaddress,twocolumn]{revtex4}
\usepackage{latexsym}
\usepackage{amsfonts}
\usepackage{amssymb}
\usepackage{amsmath}
\usepackage{graphicx}

\begin{document}

\title{Evidence for Strain-Induced Local Conductance Modulations in Single-Layer Graphene on $\rm SiO_2$}
\author{M.~L. Teague,$^{1}$ A.~P. Lai,$^{1}$ J. Velasco,$^{2}$ C.~R. Hughes,$^{1}$ A.~D. Beyer,$^{1}$ M.~W. Bockrath,$^{3}$ C.~N. Lau,$^{2}$ and N.-C. Yeh\footnote{To whom correspondence should be addressed. E-mail: ncyeh@caltech.edu}}

\affiliation{Department of Physics, California Institute of Technology, Pasadena, CA 91125\\
$^{2}$Department of Physics and Astronomy, University of California, Riverside, CA 92521\\
$^{3}$Department of Applied Physics, California Institute of Technology, Pasadena, CA 91125}

\date{\today}

\begin{abstract}
Graphene has emerged as an electronic material that is promising for device applications and for studying two-dimensional electron gases with relativistic dispersion near two Dirac points. Nonetheless, deviations from Dirac-like spectroscopy have been widely reported with varying interpretations. Here we show evidence for strain-induced spatial modulations in the local conductance of single-layer graphene on $\rm SiO_2$ substrates from scanning tunneling microscopic (STM) studies. We find that strained graphene exhibits parabolic, U-shaped conductance vs. bias voltage spectra rather than the V-shaped spectra expected for Dirac fermions, whereas V-shaped spectra are recovered in regions of relaxed graphene. Strain maps derived from the STM studies further reveal direct correlation with the local tunneling conductance. These results are attributed to a strain-induced frequency increase in the out-of-plane phonon mode that mediates the low-energy inelastic charge tunneling into graphene.
\end{abstract}

\maketitle

Since its experimental isolation in 2005,~\cite{Novoselov05} graphene has emerged as an electronic material with superior properties~\cite{Novoselov05,Novoselov07,Miao07} that are promising for device applications.~\cite{Novoselov07,Tombros07,Standley08} Additionally, graphene represents a unique opportunity to study two-dimensional electron gases~\cite{Novoselov05,Semenoff84} with relativistic dispersion near two Dirac points in the Brillouin zone.~\cite{Tombros07} That is, the charge carriers in ideal graphene are massless Dirac fermions with a conical energy-momentum relationship, $E = \pm \hbar k v_F$, where $v_F$ is the Fermi velocity, $\hbar = h/(2 \pi)$, $h$ is the Planck constant, and $k$ the momentum.~\cite{Novoselov05} Thus, the low energy charge excitations of ideal graphene are expected to obey the Dirac equation rather than the Schr$\rm \ddot{o}$dinger wave equation.~\cite{Novoselov05} In addition to the prediction of novel low-energy excitations, various interesting phenomena have been observed in bulk measurements of graphene, including the ambipolar electric field effect,~\cite{Novoselov05} the integer quantum Hall effect (IQHE),~\cite{Zhang05,Novoselov08,Matsui05,Gusynin05} the presence of a minimum conductance $4e^2/(\hbar \pi)$ in the limit of zero charge-carriers,~\cite{Miao07,Martin08} ultra high mobilities,~\cite{Du08,Bolotin08} and the absence of quantum-interference magnetoresistance.~\cite{Novoselov07,McCann06} Microscopically, while a number of investigations have been conducted on single point spectroscopy of graphene using scanning tunneling spectroscopy (STS),~\cite{Novoselov07,Zhang08,Mallet07,Li08} information on spatially resolved local density of states (LDOS) over extensive areas has been limited except for recent studies of epitaxial graphene on SiC(0001) and Ru(0001).~\cite{Rutter09,deParga08} 

In contrast to the lack of spatially resolved spectroscopic information, substantial high-resolution topographic studies of graphene have been carried out using STM. For instance, it is found that strong surface corrugations, up to 1 nm variation in height on a lateral scale of 10 nm, appear in the topography of both graphene on $\rm SiO_2$ substrates~\cite{Ishigami07,Stolyarova07} and suspended graphene.~\cite{Meyer07} These findings suggest that not only does the graphene deviate from a true two-dimensional system with a slight three-dimensional component, but the planar  structure also deviates from a perfect honeycomb lattice. Such massive changes in the topology are expected to affect the LDOS~\cite{deParga08,Cortijo07,Zhou06,Parornd06,Herbut08} as well as the phonon modes of graphene. Recent findings of out-of-plane phonon-mediated inelastic tunneling from STS stuides of mechanically exfoliated graphene~\cite{Zhang08} have revealed the important role of phonons in the low-energy excitation spectra of graphene.~\cite{Wehling08} Additionally, substrate-induced sublattice symmetry-breaking in the case of expitaxially grown graphene can give rise to energy gaps in graphene electronic structures.~\cite{KimS08} Thus, surface corrugations and structures associated with the substrate may induce significant variations in the low-energy tunneling spectra of graphene. 

In this work we address the issue of possible effects of lattice distortions on the local tunneling conductance by STM studies of single-layer graphene on $\rm SiO_2$. Graphene samples were manufactured via mechanical exfoliation. An optical microscope was used to identify the location of single-layer graphene, and photolithographic processes were employed to attach gold electrical contacts to the graphene. Prior to STM measurements, the sample was annealed in a high oxygen environment at temperatures of 400 $\rm ^{o}$C for 15 minutes to remove photoresist that remained from the photolithography processes; otherwise, atomic resolution would not have been achievable and spectroscopic measurements would have been contaminated. The topographic and spectroscopic measurements were taken with a home built cryogenic STM, which was capable of variable temperature control from room temperature to 6 K and was also compatible with magnetic fields. For studies reported in the following, we kept the measurement conditions at 77 K under high vacuum ($< 10^{-7}$ torr) and in zero magnetic field. Topographic and spectroscopic measurements were performed simultaneously at every location in a (128 $\times$ 128) pixel grid. At each pixel location, the tunnel junction was independently established so that the junction resistance of 4G$\Omega$ was maintained across the sample. The differential conductance, $(dI/dV)$, was calculated from the best polynomial fit of each current ($I$) vs. bias voltage ($V$) curve. 

Topographic STM measurements were made over numerous regions across the sample, as exemplified in Fig.~1(a) for a (2.2 nm $\times$ 5.0 nm) area. We found surface corrugations of up to $\pm 0.5$ nm over a lateral distance of 10 nm, as shown by the histogram in Fig.~1(b), which is in good agreement with the results reported previously.~\cite{Stolyarova07,Meyer07,Cortijo07} Additional confirmation of the surface corrugation was obtained from atomic force microscopy (AFM) measurements over the same region, which recorded height variations of $\pm 1.0$ nm. Further topographic STM studies over a larger area of the graphene sample revealed similar landscape of surface corrugations, as shown in Fig.~1(c). The corresponding histogram of the height variations is depicted in Fig.~1(d).  

\begin{figure}
\includegraphics[width=3.4in]{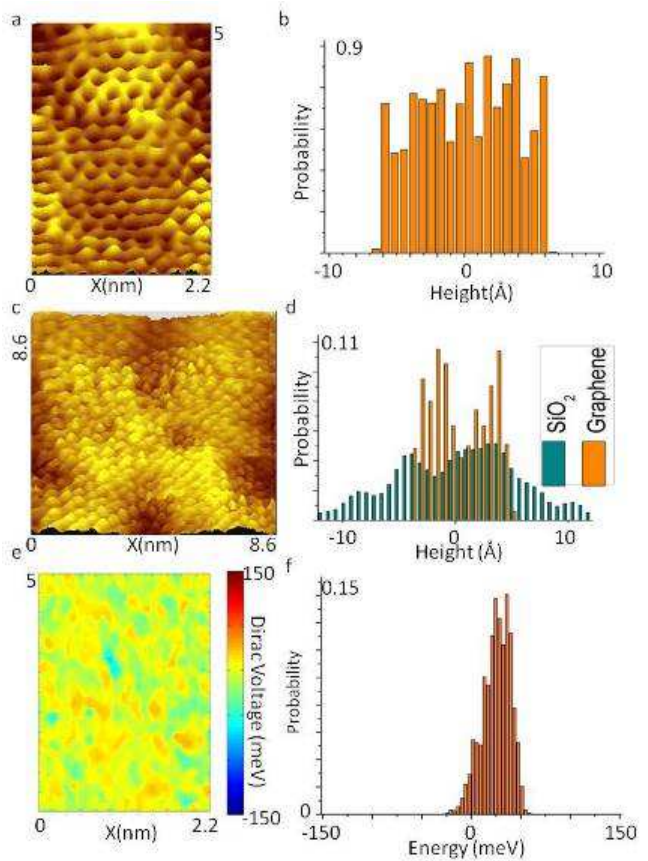}
\caption{Surface corrugations of graphene on $\rm SiO_2$ substrate: (a) Topographic image of graphene surface over a (2.2 nm $\times$ 5.0 nm) area. (b) Histogram of the graphene height variations over the area shown in Fig.~1(a). (c) Topographic image of graphene surface over a larger (8.6 nm $\times$ 8.6 nm) area. (d) Histogram of the graphene height variations over the larger area shown in Fig.~1(c) is compared with the histogram of a bare SiO$_2$ substrate over the same area, showing apparent correlation between the height variations of graphene and those of the underlying SiO$_2$ substrate. (e) Two-dimensional map of the Dirac voltage over the area in Fig.~1(a), showing relatively homogeneous Dirac voltage distributions. (f) Histogram of the Dirac voltage over the same area in Fig.~1(e).}
\label{Fig1}
\end{figure}

To understand the physical origin for the graphene surface corrugations, we peformed AFM measurements on a bare $\rm SiO_2$ substrate, which yielded height variations of $\pm 2.0$ nm. In Fig.~1(d) we compare the histogram of the graphene surface corrugations with that of the $\rm SiO_2$ substrate over a (8.6 nm $\times$ 8.6 nm) area. The apparent correlations between the two histograms suggest that the surface corrugations of the graphene sample result from the underlying roughness of the substrate. We further investigated the possibility of graphene surface roughness resulting from the gold contacts on the graphene sample, and found no apparent correlation between the surface roughness and the distance from the gold contacts. Additionally, extensive topographic surveys revealed no discernible atomic defects in our graphene sample. 

\begin{figure}
\includegraphics[width=3.0in]{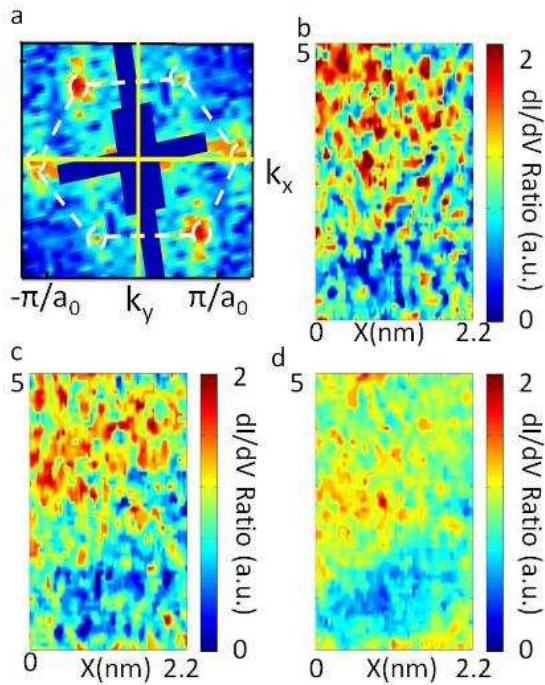}
\caption{Correlation of spatially varying topography and conductance maps of graphene: (a) FFT of the topographic map in Fig.~1(a), showing a distorted hexagon. (b) A conductance ratio map taken at $V_B = 0$, revealing significant spatial variations in the conductance that are correlated with the topography in Fig.~1(a). Here the conductance ratio is defined as the local conductance relative to the mean conductance of the entire area in the field of view. (c) A conductance ratio map taken at $V_B = 80$ meV, showing weaker spatial variations in the conductance relative to Fig.~2(b). (d) A conductance ratio map taken at $V_B = 240$ meV, showing reduced spatial variations in the conductance relative to Figs.~2(b) and 2(c).}
\label{Fig2}
\end{figure}

To quantify the lattice distortions of graphene on $\rm SiO_2$, we performed fast Fourier transformation (FFT) of the topographical scan over the (2.2 nm $\times$ 5.0 nm) region in Fig.~1(a). The FFT in Fig.~2(a) revealed a distorted hexagon, indicating significant deformation in the lattice structure. Thus, spatially varying strain maps may be derived from studying the displacement fields $\textbf{u}$ of local regions. Specifically, we may choose one of the principle lattice vectors as the $x$-axis, and define the local displacement field $\textbf{u}(x,y) \equiv u_x \hat{x} + u_y \hat{y}$ as the difference of the local lattice vector from the equilibrium lattice vector. Following similar approaches previously developed for electron microscopic studies of other materials,~\cite{Hytch03} we obtained the strain maps shown in Figs.~3(a)-3(d) for the sample area given in Fig.~1a: Fig.~3(a) illustrates the ``scalar strain'' map $S_0 (x,y) \equiv |\textbf{u}(x,y)/a_0|$, and Figs.~3(b)-3(d) correspond to the maps of the strain tensor components $S_{xx}(x,y) \equiv (\partial u_x/\partial x)$, $S_{xy}(x,y) = S_{yx}(x,y) \equiv \lbrack (\partial u_x/\partial y) + (\partial u_y/\partial x) \rbrack /2$, and $S_{yy}(x,y) \equiv (\partial u_y/\partial y)$. From the strain maps we find that on the whole the upper section of the graphene in view was more strained while the lower section was more relaxed.  
\begin{figure}
\includegraphics[width=3.0in]{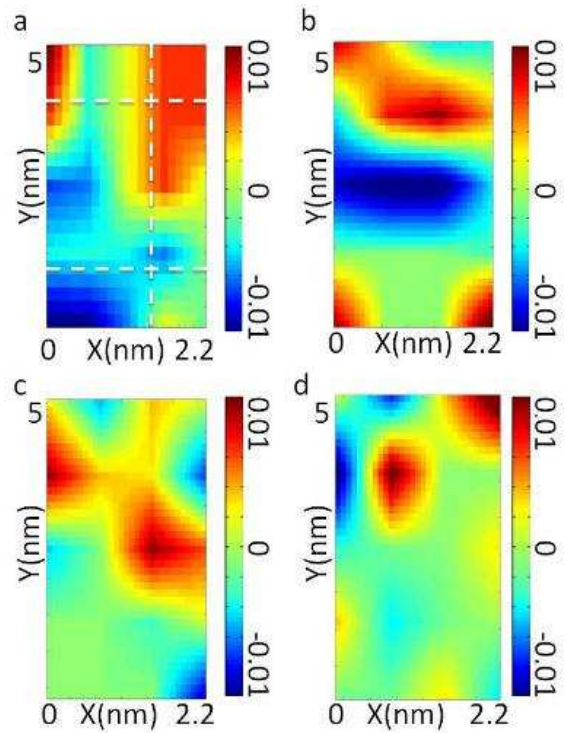}
\caption{Strain maps of graphene over a (2.2 nm $\times$ 5.0 nm) area: (a) A ``scalar strain'' map $S_0 (x,y)$, showing the spatially varying fraction of lattice distortion relative to the equilibrium lattice constant $a_0 = 2.46$\AA. The three white dashed lines (upper, lower and vertical) indicate three different line-cuts of varying strain tensor components. (See Fig.~4.) (b) A map for the strain tensor component $S_{xx} (x,y)$. (c) A map for the strain tensor component $S_{xy} (x,y)$. (d) A map for the strain tensor component $S_{yy} (x,y)$. }
\label{Fig3}
\end{figure}

\begin{figure}
\includegraphics[width=3.5in]{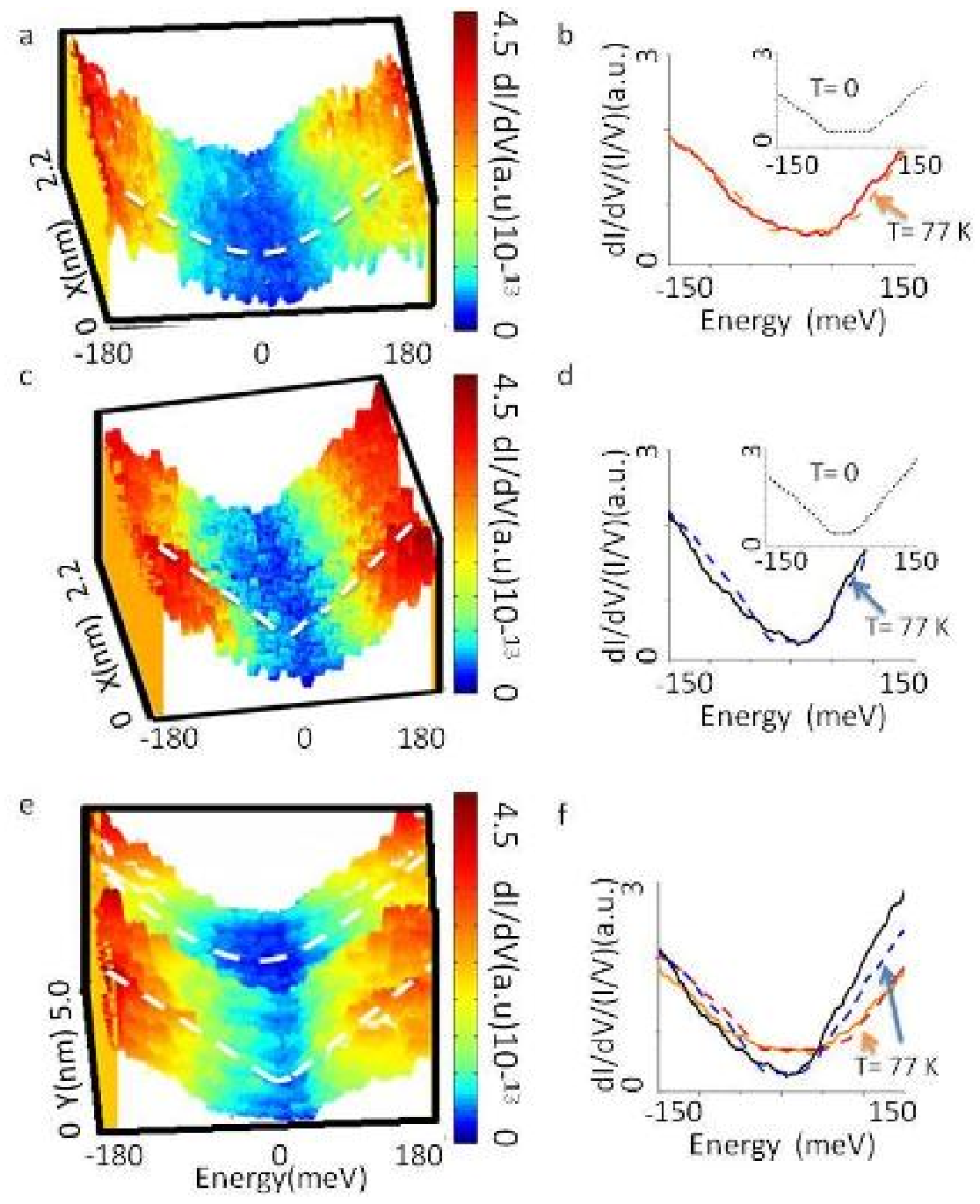}
\caption{Correlation of the tunneling spectral characteristics at $T = 77$ K with the strain tensor components: (a) A series of differential conductance ($dI/dV$) vs. bias voltage ($V_B$) spectra along the upper line-cut in Fig.~3(a), showing primarily parabolic U-shaped spectra around $V_B = 0$. (b) A representative U-shaped normalized conductance $(dI/dV)/(I/V)$ spectrum (solid curve) in the strained region. A theoretical curve with an out-of-plane phonon frequency $\hbar \omega _a = 44$ meV ~\cite{Wehling08} is shown by the dashed line. The inset shows the theoretical curves at $T = 0$. (c) A series of ($dI/dV$)-vs.-$V_B$ spectra along the bottom line-cut in Fig.~1(a), showing primarily Dirac-like V-shaped spectra around $V_B = 0$. (d) A representative V-shaped normalized conductance spectrum in the relaxed region together with the theoretical fitting curve (dashed line) with a phonon frequency $\hbar \omega _a = 26$ meV. The theoretical curve at $T = 0$ is illustrated in the inset. (e) A series of ($dI/dV$)-vs.-$V_B$ spectra along the vertical line-cut in Fig.~3(a), showing evolution from primarily U-shaped spectra to V-shaped spectra. (f) Comparison of two representative normalized conductance spectra from strained to relaxed regions. The theoretical fitting curves correspond to a higher phonon frequency $\hbar \omega _a = 41$ meV associated with the strained region and a smaller frequency $\hbar \omega _a = 24$ meV with the relaxed region.}
\label{Fig4}
\end{figure}

In addition to the togographic images, spatially resolved spectroscopic studies were carried out across the graphene sample to investigate possible correlation between the surface corrugations and modulations of the LDOS. The differential conductance $(dI/dV)$ as a function of bias voltage $(V)$ was obtained at each pixel of the sample area under investigation. We first noticed that at each pixel the corresponding spectrum always exhibited a minimum conductance at a finite bias voltage $V_D$. This finite $V_D$ value, known as the Dirac voltage, corresponds to an offset in Fermi energy, and the predominant mechanism for this offset may be attributed to charge impurities on the $\rm SiO_2$ surface. For the area shown in Fig.~1(a), the corresponding two-dimensional map and the histogram of the Dirac voltage $V_D (x,y)$ are illustrated in Figs.~1(e) and 1(f), respectively, showing relatively uniform variations without apparent reference to the topography in Fig.~1(a). In contrast, the spatial map of tunneling conductance at $V_B \equiv (V - V_D) = 0$, Fig.~2(b), reveals strong correlation with the topography in Fig.~1(a). As $V_B$ increases, the spatial variations in conductance across the sample decrease steadily, as exemplified by Fig.~2(c) for $V_B = 80$ meV and Fig.~2(d) for $V_B = 240$ meV. The lack of correlation between the $V_D (x,y)$ map in Fig.~1(e) and the low-energy conductance map in Fig.~2(b) suggests that charge impurities are unlikely the primary cause for the spatial modulations in the LDOS. 

To understand possible correlations between the spatial modulations of tunneling conductance and the local strain, we further examined the tunneling spectra and the strain tensor components along three different line-cuts (upper, vertical, lower) across the sample, with the three line-cuts indicated by the white dashed lines in Fig.~3(a). The spatial evolution of the spectra associated with the upper line-cut is given in Fig.~4(a), which are mostly parabolic (U-shaped) around $V_B = 0$, as exemplified by the solid curve in Fig.~4(b). We further note that the spatial distribution of these U-shaped spectra correlates well with the region of strained graphene according to the maps in Figs.~3(b)-3(d). 

If we employ the scenario of out-of-plane phonon-mediated inelastic tunneling~\cite{Wehling08} by allowing the out-of-plane phonon frequency $\omega _a$ as a fitting parameter and taking $T = 77$ K, as elaborated in the next paragraph, we find that the theoretical fitting (dashed curve in Fig.~4(b)) agrees reasonably well with the experimental spectra for a phonon frequency $\hbar \omega _a = 44$ meV. In contrast, the spectra of the lower line-cut in Fig.~4(c) are more Dirac-like (V-shaped) around $V_B = 0$, as exemplified by the solid curve in Fig.~4(d). Similarly, the region where such V-shaped spectra dominate coincides with the area of relaxed graphene with nearly vanishing strain tensor components as shown in Figs.~3(b)-3(d). Moreover, the theoretical fitting to the spectrum in Fig.~4(d) yields a smaller out-of-plane phonon frequency $\hbar \omega _a = 26$ meV relative to that in the strained region. For the vertical line-cut that moves from a region of strained to relaxed graphene, the corresponding spectra in Fig.~4(e) exhibits evolution from spectra with larger $\hbar \omega _a$ values to those with smaller $\hbar \omega _a$ values while the strain tensor components decrease according to Figs.~3(b)-3(d). Direct comparison of two representative spectra from strained to relaxed regions with a decreasing out-of-plane phonon frequency is shown in Fig.~4(f). The apparent correlation of spectral characteristics with the strain maps together with their lack of correlation with the Diract voltage $V_D(x,y)$ suggest that strain-induced conductance modulations and phonon-mediated inelastic tunneling rather than charge impurities are the primary cause for spectral deviations from Dirac behavior in graphene.

Next, we specify our fitting to the tunneling conductance in Fig.~4 based on the three-band model that considers the mixing of nearly-free electron bands at the zone center $\Gamma$ with the Dirac fermions at the zone edge K and K$^{\prime}$ through coupling with the out-of-plane phonons.~\cite{Wehling08} According to this scenario, the tunneling DOS from STM measurements is dominated by the DOS of the nearly-free electron bands, $N_{\Gamma}(\omega)$. In the limit of $|\omega| \ll W$ where $W \approx 6$eV denotes the Dirac electron bandwidth, $N_{\Gamma}(\omega)$ in the absence of any gate voltage ($\mu = 0$) is given by:~\cite{Wehling08} 
\begin{equation}
\label{eq1}
N_{\Gamma} (\omega) = - \frac{\Sigma_{1,1}^{\prime \prime} (\omega + i \delta)}{\pi |\omega - E_{\sigma} - \Sigma_{1,1} (\omega + i \delta)|^2}.
\end{equation}
Here $E_{\sigma} \approx 3.3$eV, and the real and imaginary parts of the electron self-energy $\Sigma_{1,1} (\omega + i \delta) = \Sigma_{1,1} ^{\prime} (\omega + i \delta) + i \Sigma_{1,1} ^{\prime \prime} (\omega + i \delta)$ satisfy the following energy dependence:~\cite{Wehling08}   
\begin{eqnarray}
\label{eq2}
\Sigma_{1,1} ^{\prime} (\omega + i \delta) &\propto (\omega - \omega _a) \log |\frac{\omega - \omega _a}{W}| + (\omega + \omega _a) \log |\frac{\omega + \omega _a}{W}|  \quad \cr
\Sigma_{1,1} ^{\prime \prime} (\omega + i \delta) &\propto - \Theta (|\omega| - \omega _a) |\omega - {\rm sgn} (\omega) \omega _a|. \qquad \qquad \qquad
\end{eqnarray}
In our data analysis we assume $\omega _a$ as a fitting parameter, and also allow a small constant zero-bias offset in Eq.~(1), which may be attributed to enhanced LDOS due to impurities.~\cite{Zhang08,Wehling08} For $T =$ 77 K, we replace $\omega$ by $i \omega _n$ and sum over the fermion Matsubara frequencies $\omega _n$ with a prefactor $(1/\beta)$ where $\beta ^{-1} = 77$ K. Given that the thermal energy at 77 K is much smaller than $\hbar \omega _a$, however, the thermal smearing effect is insignificant, as manifested by Figs.~4(b) and 4(d).

Finally, we note that the $\hbar \omega _a$ values derived from the phonon-mediated tunneling scenario appear to vary significantly not only within different regions of a substrate but even more among different substrates, ranging from $24 \sim 44$ meV found in our experiments to 60 meV and 100 meV reported in Refs.~\cite{Zhang08} and \cite{Mallet07}, respectively. In the case of graphene grown on SiC substrates,~\cite{Mallet07} superlattice modulations are known to have induced modifications to the electronic bandstructures,~\cite{KimS08} which may account in part for the significantly larger gap features. Although we cannot pinpoint the exact origin of gap variations for gaphene prepared under different conditions, the significant strain-induced effects appear to be at least partially responsible for the differences. On the other hand, it seems unlikely for the out-of-plane phonon mode alone to account for the significant variations of the empirical $\hbar \omega _a$ values among the tunneling spectra of graphene on different substrates. Further investigation appears necessary to address this issue. 
      
In summary, our findings of substrate-induced lattice strains in graphene and the apparent correlation between the strain tensor components with modulations in the local tunneling conductance provide new insights into mechanisms that can strongly influence the low-energy excitations of Dirac fermions. From the viewpoint of device applications of graphene, our studies suggest that strain-induced variations in the electronic properties are limited to relative low excitation energies, around a few tens of meV's. Therefore, device applications involving finite gate or bias voltages on the order of eV's~\cite{Novoselov07,Tombros07,Standley08} will not be sensitive to the substrate-induced strain effect. On the other hand, it is conceivable to apply strain-induced effects to controlling the low-bias electronic properties of graphene-based devices. In this context, our findings of strain-induced energy gap variations in graphene can be of technological relevance. 

\begin{acknowledgments}
The work at Caltech was jointly supported by NSF and NRI under the Center of Science and Engineering of Materials (CSEM). The work at University of California, Riverside was supported by NSF CAREER DMR/0748910. We thank Dr. A. V. Balatsky for useful discussions.
\end{acknowledgments}

\bibliographystyle{apsrev}

\begin{thebibliography}{30}

\bibitem{Novoselov05}
Novoselov, K. S.; Geim, A. K.; Morozov, S. V.; Jiang, D.; Katsnelson, M. I.; Grigorieva, I. V.; Dubonos, S. V.; Firsov, A. A. {\it Nature} {\bf 2005}, {\it 438}, 197-200.

\bibitem{Novoselov07}
Geim, A. K.; Novoselov, K. S. {\it Nature Materials} {\bf 2007}, {\it 6}, 183-191

\bibitem{Miao07}
Miao, F.; Wijeratne, S.; Zhang, Y.; Coskun, U. C.; Bao, W.; Lau, C. N. {\it Science} {\bf 2007}, {\it 317}, 1530-1533.

\bibitem{Tombros07}
Tombros, N.; Jozsa, C.; Popinciuc, M.; Jonkman, H. T.; van Wees, B. J.  {\it Nature} {\bf 2007}, {\it 448}, 571-574.

\bibitem{Standley08} 
Standley, B.; Bao, W. Z.; Zhang, H.; Bruck, J.; Lau C. N.; Bockrath, M. W. {\it Nano Lett.} {\bf 2008}, {\it 8}, 3345-3349.

\bibitem{Semenoff84}
Semenoff, G. W. {\it Phys. Rev. Lett.} {\bf 1984}, {\it 53}, 2449-2452.

\bibitem{Zhang05}
Zhang Y. B.; Tan, Y. W.; Stormer, H. L.; Kim, P.  {\it Nature} {\bf 2005}, {\it 438}, 201-204.

\bibitem{Novoselov08}
Novoselov, K. S.; Jiang, Z.; Zhang, Y.; Morozov, S. V.; Stormer, H. L.; Zeitler, U.; Maan, J. C.; Boebinger, G. S.; Kim, P.; Geim, A. K. {\it Science} {\bf 2007}, {\it 315}, 1379.

\bibitem{Matsui05}
Matsui, T.; Kambara, H.; Niimi, Y.; Tagami, K.; Tsukada, M.; Fukuyama, H. {\it Phy. Rev. Lett.} {\bf 2005}, {\it 94}, 226403.

\bibitem{Gusynin05}
Gusynin, V. P.; Sharapov, S. G. {\it Phys Rev. Lett.} {\bf 2005}, {\it 95}, 146801.

\bibitem{Martin08}
Martin, J.; Akerman, N.; Ulbricht, G.; Lohmann, T.; Smet, J. H.; Von Klitzing, K.; Yacoby, A. {\it Nature Physics} {\bf 2008}, {\it 4}, 144-148.

\bibitem{Du08}
Du, X.; Skachko, I.; Barker, A.; Andrei, E. Y. {\it Nature Nanotechnology} {\bf 2008}, {\it 3}, 491 - 495.

\bibitem{Bolotin08}
Bolotin, K. I.; Sikes, K. J.; Jiang, Z.; Klima, M.; Fudenberg, G.; Hone, J.; Kim, P.; Stormer, H. L. {\it Solid State Commun.} {\bf 2008}, {\it 146}, 351-355.

\bibitem{McCann06}
McCann, E.; Kechedzhi, K.; Fal'ko, V. I.; Suzuura, H.; Ando, T.; Altshuler, B. L.; {\it Phys Rev. Lett.} {\bf 2006}, {\it 97}, 146805.

\bibitem{Zhang08}
Zhang, Y. B.; Brar, V. W.; Wang, F.; Girit, C.; Yayon, Y.; Panlasigui, M.; Zettl, A.; Crommie, M. F. {\it Nature Physics} {\bf 2008}, {\it 4}, 627-630.

\bibitem{Mallet07}
Mallet, P.; Varchon, F.; Naud, C.; Magaud, L.; Berger, C.; Veuillen, J.-Y. {\it Phys. Rev. B} {\bf 2007}, {\it 76}, 041403.

\bibitem{Li08}
Li, G. {\it et al.} {\it arXiv:} 0803.4016 {\bf 2008}.

\bibitem{Rutter09}
Rutter, G. M.; Crain, J. N.; Guisinger, N. P.; Li, T.; First, P. N.; Stroscio, J. A. {\it Science} {\bf 2009}, {\it 317}, 219-222.

\bibitem{deParga08}
de Parga, A. L. V.; Calleja, F.; Borca, B.; Passeggi, M. C. G. Jr.; Hinarejos, J. J.; Guinea, F.; Miranda, R. {\it Phys. Rev. Lett.} {\bf 2008}, {\it 100}, 056807. 

\bibitem{Ishigami07}
Ishigami, M.; Chen, J. H.; Cullen, W. G.; Fuhrer, M. S.; Williams, E. D. {\it Nano Letters} {\bf 2007}, {\it 7}, 1643-1648.

\bibitem{Stolyarova07}
Stolyarova, E.; Rim, K. T.; Ryu, S.; Maultzsch, J.; Kim, P.; Brus, L. E.; Heinz, T. F.; Hybertsen, M. S.; Flynn, G. W. {\it Proc. Natl. Acad. Sci.} {\bf 2007}, {\it 104}, 9209-9212.

\bibitem{Meyer07}
Meyer, J. C.; Geim, A. K.; Katsnelson, M. I.; Novoselov, K. S.; Booth, T. J.; Roth, S. {\it Nature} {\bf 2007}, {\it 446}, 60-63.

\bibitem{Cortijo07}
Cortijo, A.; Vomediano, M. A. H. {\it Nucl. Phys. B} {\bf 2007}, {\it 763}, 293-308.

\bibitem{Zhou06}
Zhou, S. Y.; Gweon, G.-H.; Lanzara, A. {\it Annals of Physics} {\bf 2006}, {\it 321}, 1730-1746. 

\bibitem{Parornd06}
Parornd, B.; Peeters, F. M. {\it Phys. Rev. B} {\bf 2006}, {\it 74}, 075404.

\bibitem{Herbut08}
Herbut, I. F.; Juricic, V.; Vafek, O. {\it Phys. Rev. Lett.} {\bf 2008}, {\it 100}, 046403. 

\bibitem{Wehling08}
Wehling, T. O.; Grigorenko, I.; Lichtenstein, A. I.; Balatsky, A. V. {\it Phys. Rev. Lett.} {\bf 2008}, {\it 101}, 216803.

\bibitem{KimS08}
Kim, S.; Ihm, J.; Choi, H. J.; and Son, Y.-W. {\it Phys. Rev. Lett.} {\bf 2008}, {\it 100}, 176802.

\bibitem{Hytch03}
H$\rm \ddot{y}$tch, M. J.; Putaux, J. L.; Pénisson, J. M. {\it Nature} {\bf 2003}, {\it 423}, 270–273.

\end{thebibliography}

\end{document}